\documentclass[epj]{webofc}\usepackage[varg]{txfonts}
\usepackage{bm}\woctitle{QCD@Work 2014}\begin{document}
\title{Decay Constants of Beauty Mesons from QCD Sum
Rules}\author{Wolfgang Lucha\inst{1}\fnsep
\thanks{\email{Wolfgang.Lucha@oeaw.ac.at}}\and Dmitri Melikhov
\inst{2,3}\fnsep\thanks{\email{dmitri_melikhov@gmx.de}}\and
Silvano Simula\inst{4}\fnsep\thanks{\email{simula@roma3.infn.it}}}
\institute{Institute for High Energy Physics, Austrian Academy of
Sciences, Nikolsdorfergasse 18, A-1050 Vienna, Austria\and Faculty
of Physics, University of Vienna, Boltzmanngasse 5, A-1090 Vienna,
Austria\and D.~V.~Skobeltsyn Institute of Nuclear Physics,
M.~V.~Lomonosov Moscow State University, 119991, Moscow,
Russia\and INFN, Sezione di Roma Tre, Via della Vasca Navale 84,
I-00146, Roma, Italy}\abstract{Our recently completed analysis of
the decay constants of both pseudoscalar and vector beauty mesons
reveals that in the bottom-quark sector two specific features~of
the sum-rule predictions show up: (i) For the input value of the
bottom-quark mass in the $\overline{\rm MS}$ scheme
$\overline{m}_b(\overline{m}_b)\approx4.18\;\mbox{GeV},$ the
sum-rule result $f_B\approx210$--$220\;\mbox{MeV}$ for the
$B$~meson decay constant is substantially larger than the recent
lattice-QCD finding $f_B\approx190\;\mbox{MeV}.$ Requiring QCD sum
rules to reproduce the lattice-QCD value of $f_B$ yields a
significantly larger $b$-quark mass:
$\overline{m}_b(\overline{m}_b)=4.247\;\mbox{GeV}.$ (ii) Whereas
QCD sum-rule predictions~for the charmed-meson decay constants
$f_D,$ $f_{D_s},$ $f_{D^*}$ and $f_{D_s^*}$ are practically
independent~of~the choice of renormalization scale, in the beauty
sector the results for the decay constants---and especially for
the ratio $f_{B^*}/f_B$---prove to be very sensitive to the
specific scale setting.}\maketitle

\section{Correlator, operator product expansion, and heavy-quark
mass schemes}The starting point of our QCD sum-rule evaluation of
the decay constants \cite{SVZ} of beauty mesons is the
time-ordered product of two meson interpolating currents,
\emph{viz.}, $j_5(x)=(m_b+m)\,\bar q(x)\,{\rm i}\,\gamma_5\,b(x)$
for~the~$B$ meson and $j_\mu(x)=\bar q(x)\,\gamma_\mu\,b(x)$ for
the $B^*$ meson. The correlator of pseudoscalar currents is
defined~by
$$\Pi(p^2)\equiv{\rm i}\int{\rm d}^4x\,{\rm e}^{{\rm i}\,p\,x}
\left\langle0\left|T\!\left(j_5(x)\,j^\dag_5(0)\right)\right|0
\right\rangle.$$
The Borel transform of this correlation function depends on some
Borel parameter $\tau$ and takes the~form
$$\Pi(\tau)=f_B^2\,M_B^4\exp\left(-M_B^2\,\tau\right)+
\int\limits_{(M_{B^*}+M_P)^2}^\infty{\rm d}s\,{\rm e}^{-s\,\tau}\,
\rho_{\rm hadr}(s)=\int\limits_{(m_b+m)^2}^\infty{\rm d}s\,{\rm
e}^{-s\,\tau}\,\rho_{\rm pert}(s,\mu)+\Pi_{\rm power}(\tau,\mu)\
.$$
The $B$-meson decay constant $f_B$ is defined by
$\langle0|j_5(0)|B\rangle=f_B\,M_B^2.$ In order to remove all
excited-state contributions, we adopt the standard assumption of
quark--hadron duality: the contributions of excited states are
compensated by the perturbative contribution above an
\emph{effective continuum threshold\/} $s_{\rm eff}(\tau)$ which
\emph{differs\/} from the physical continuum threshold. Applying
this Ansatz leads, for the $B$ meson,~to
$$f_B^2\,M_B^4\exp\left(-M_B^2\,\tau\right)
=\int\limits_{(m_b+m)^2}^{s_{\rm eff}(\tau)}{\rm d}s\,{\rm
e}^{-s\,\tau}\,\rho_{\rm pert}(s,\mu)+\Pi_{\rm
power}(\tau,\mu)\equiv\Pi_{\rm dual}(\tau,s_{\rm eff}(\tau))\ .$$
The right-hand side of the above relation constitutes what we call
the \emph{dual correlator\/} $\Pi_{\rm dual}(\tau,s_{\rm
eff}(\tau))$. The best-known three-loop calculation of the
perturbative spectral density $\rho_{\rm pert}$ has been performed
as an expansion in terms of the $\overline{\rm MS}$ strong
coupling, $a(\mu)\equiv\alpha_{\rm s}(\mu)/\pi$ and the $b$-quark
pole mass $M_b$~\cite{SDa,SDb}:
$$\rho_{\rm pert}(s,\mu)=\rho^{(0)}(s,M_b^2)
+a(\mu)\,\rho^{(1)}(s,M_b^2)+a^2(\mu)\,\rho^{(2)}(s,M_b^2,\mu)
+\cdots\ .$$
An tantalizing feature of the pole-mass operator product expansion
(OPE) is that each of the known perturbative contributions to the
dual correlator is positive. Unfortunately, such a pole-mass OPE
does not provide a visible hierarchy of the perturbative
contributions and thus poses strong doubts that the $O(\alpha_{\rm
s}^2)$-truncated pole-mass OPE can provide reliable estimates of
the decay constants. An alternative \cite{JL} is to reorganize the
perturbative expansion in terms of the running $\overline{\rm MS}$
mass, $\overline{m}_b(\nu)$,~by substituting in the spectral
densities $\rho^{(i)}(s,M_b^2)$ $M_b$ by its perturbative
expansion in terms of the running mass $\overline{m}_b(\nu)$
$$M_b=\overline{m}_b(\nu)\left[1+a(\nu)\,r_1
+a^2(\nu)\,r_2+\cdots\right].$$
The spectral densities in the $\overline{\rm MS}$ scheme are
obtained by expanding the pole-mass spectral densities~in powers
of $a(\mu)$ and omitting terms of order $O(a^3)$ and higher;
starting with order $O(a)$ they contain~two parts: the result of
Refs.~\cite{SDa,SDb} and the impact of lower perturbative orders
arising upon expansion of the pole mass in terms of the running
mass. In this way, because of the truncation of the perturbative
series, one gets an explicit \emph{unphysical\/} dependence of
both dual correlator and extracted decay constant on the scale
$\mu$. In principle, any scale should be equally good. In
practice, however, the hierarchy of perturbative contributions to
the dual correlator depends on the precise choice of the scale,
opening~the possibility to choose the scale $\mu$ such that the
hierarchy of the new perturbative expansion is improved.

\begin{figure}[t]\centering\begin{tabular}{cc}
\includegraphics[width=6.687cm]{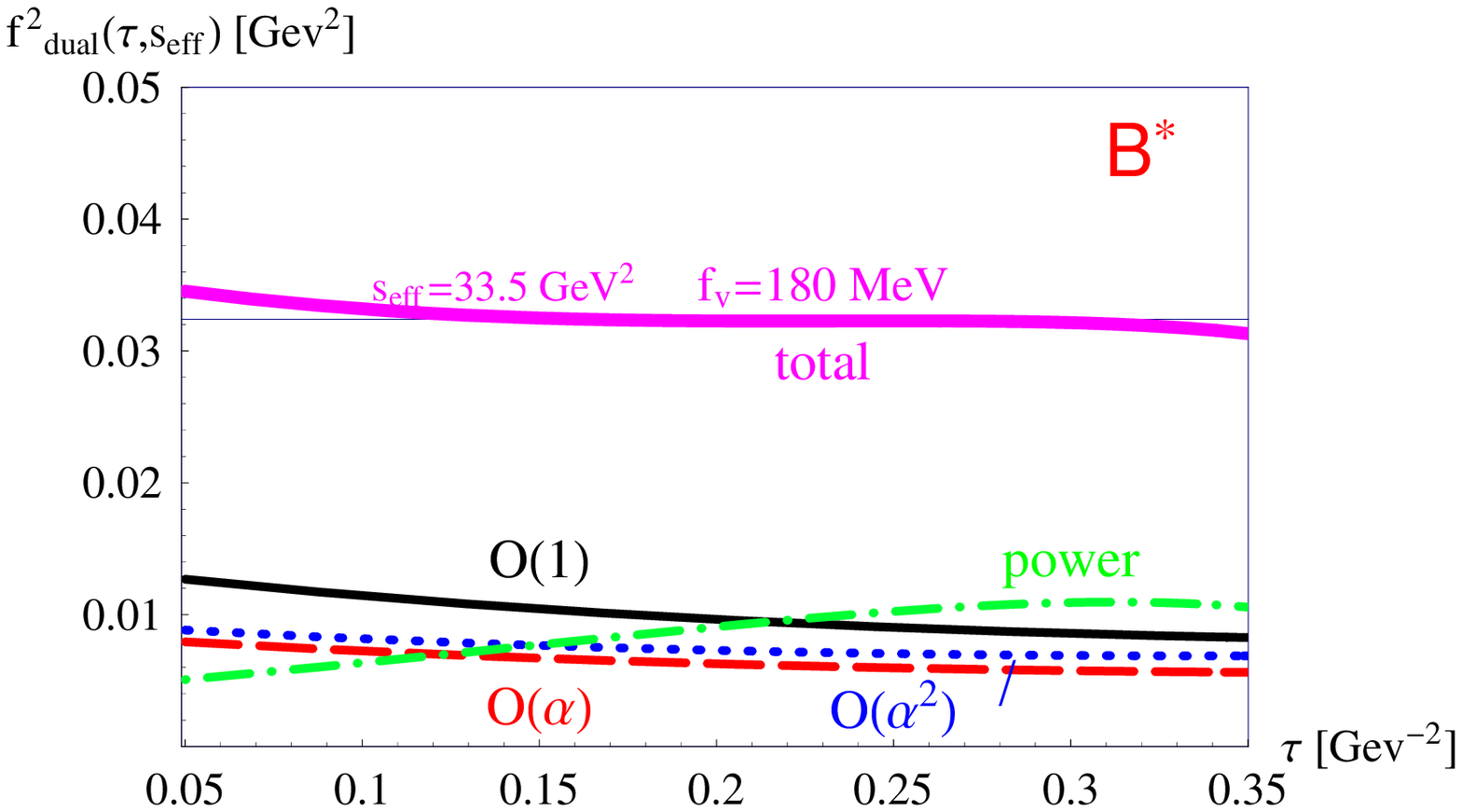}&
\includegraphics[width=6.687cm]{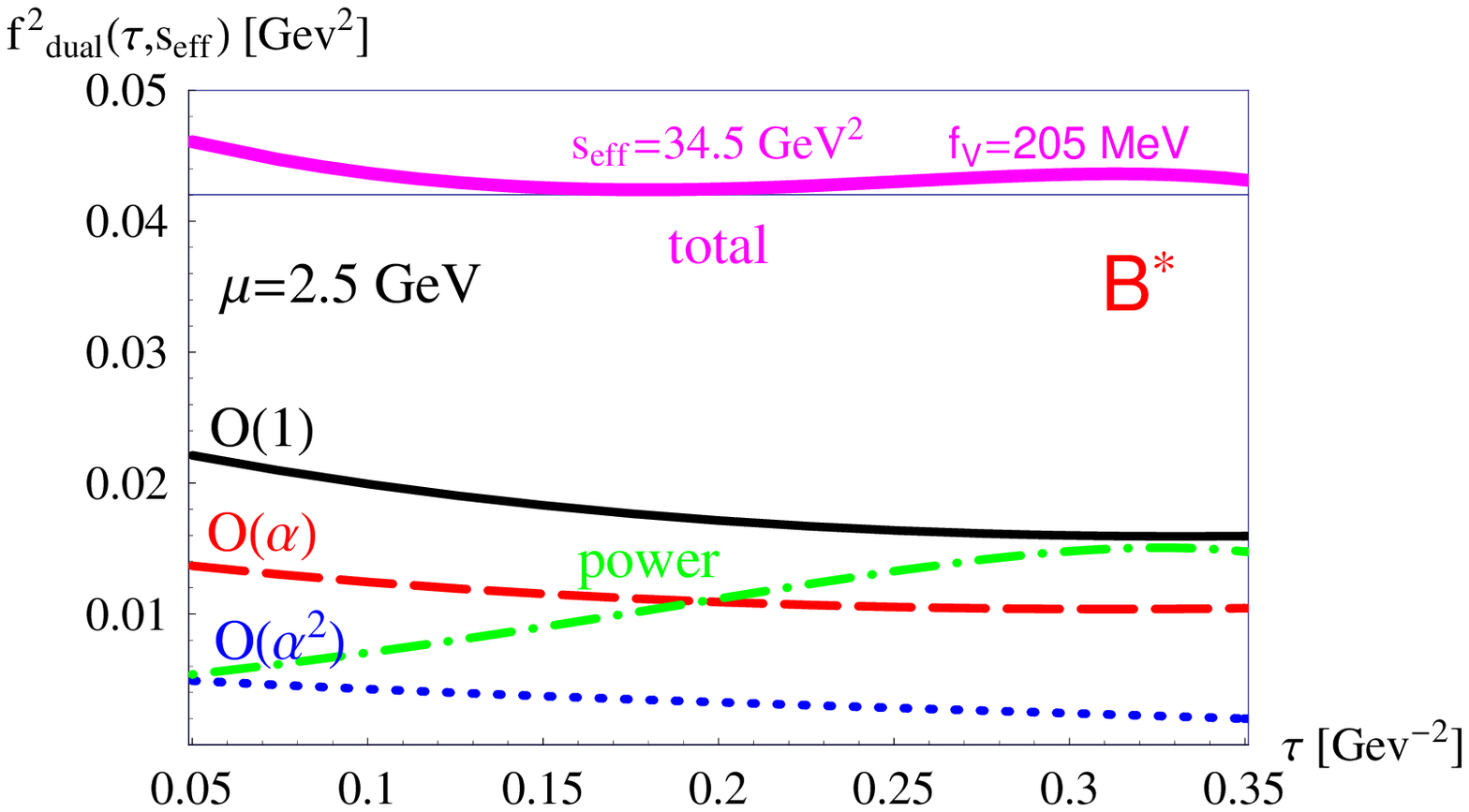}\\(a)&(b)\\
\includegraphics[width=6.687cm]{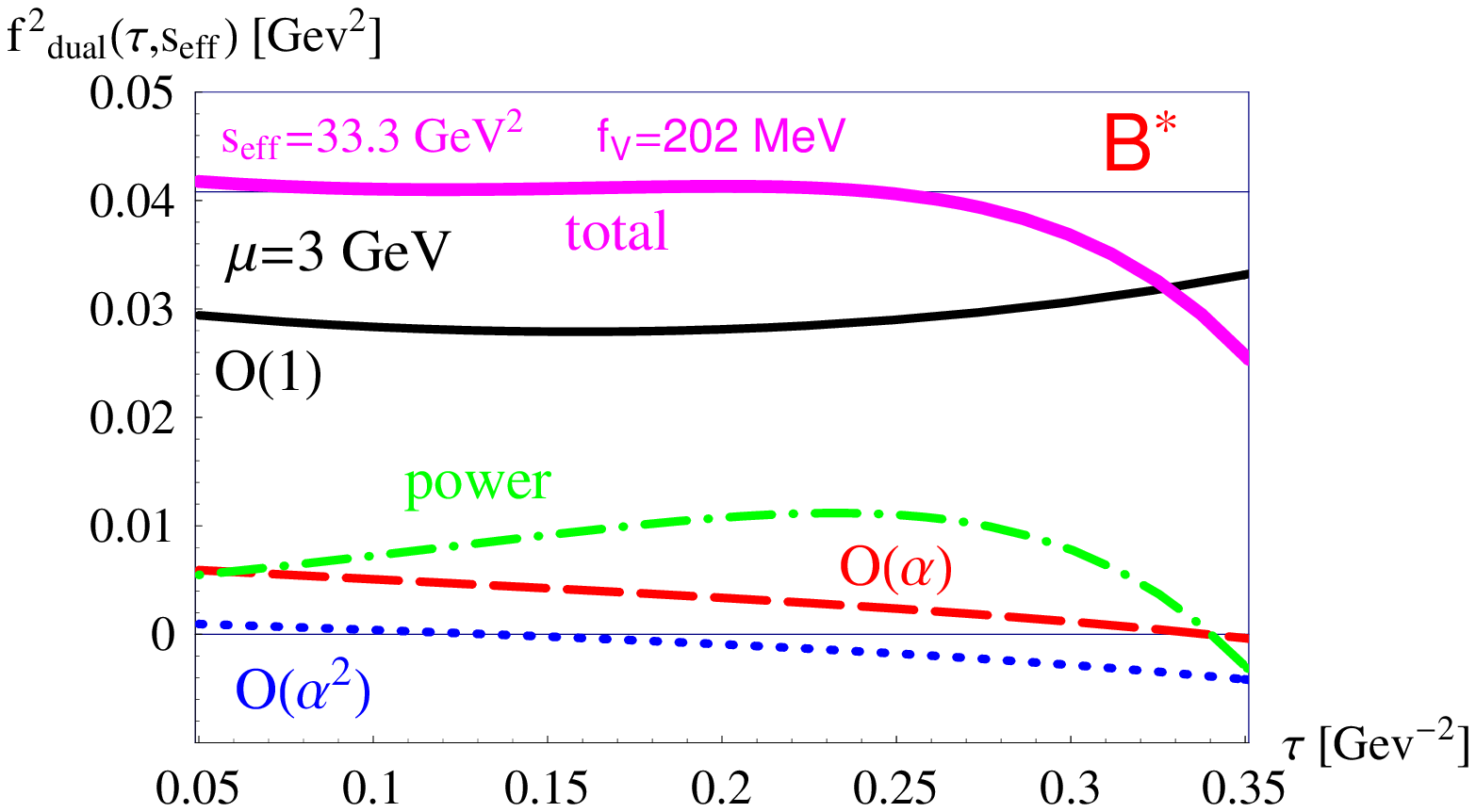}&
\includegraphics[width=6.687cm]{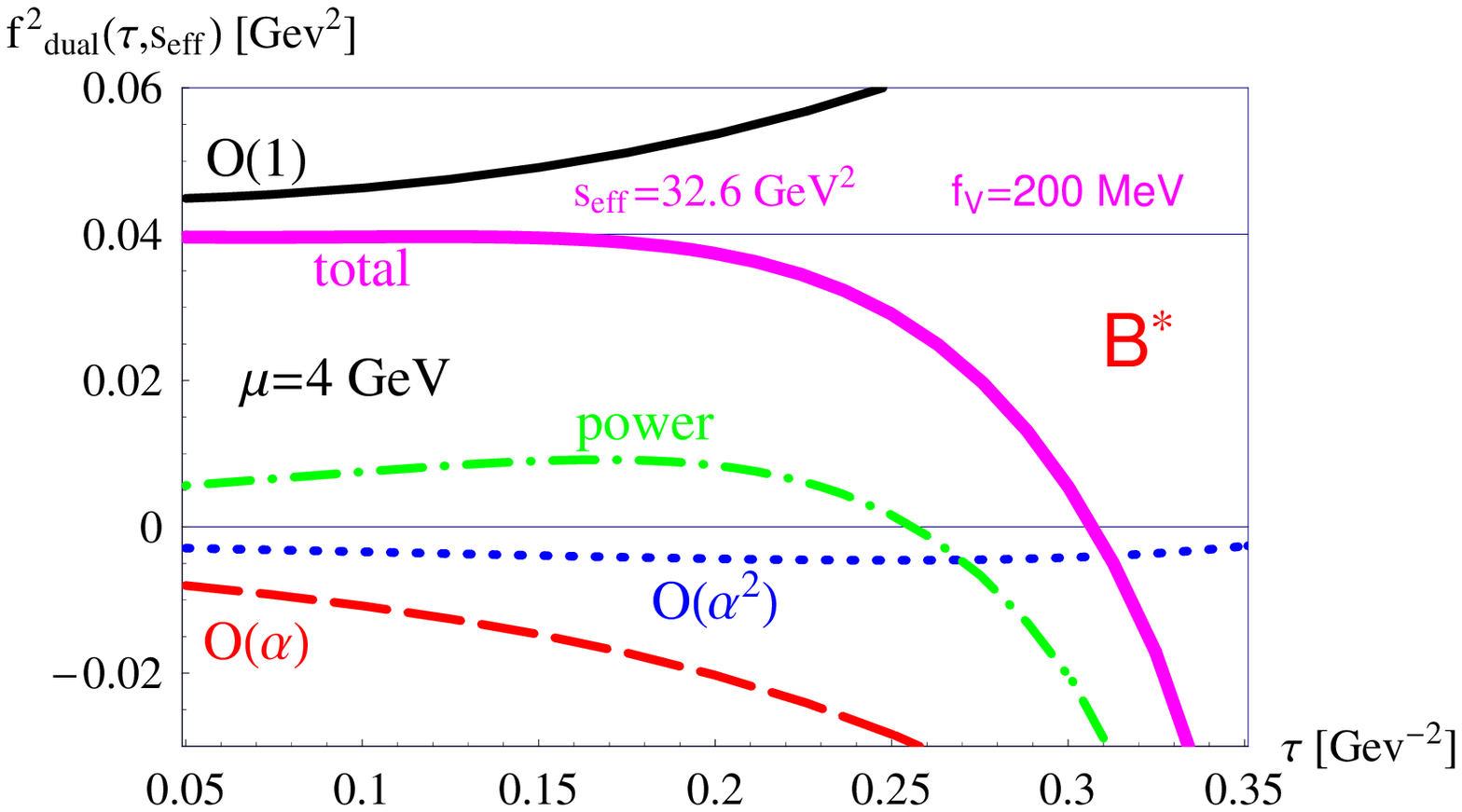}\\(c)&(d)\\
\end{tabular}\caption{QCD sum-rule estimates for the $B^*$ meson
using either pole-mass or running-mass scheme at different scales:
the $O(\alpha_{\rm s}^2)$-truncated pole-mass OPE shows no
hierarchy of the perturbative expansion and cannot be used. Even
the hierarchy of the running-mass OPE is not automatic but depends
strongly on the renormalization scale~$\mu$. For
$\overline{m}_b(\overline{m}_b)=4.18\;\mbox{GeV},$ the two-loop
pole mass is $M_b=4.80\;\mbox{GeV}.$ For each case, a constant
effective~threshold $s_{\rm eff}$ is determined by requiring
maximal stability of the predictions in the Borel window
$0.05\le\tau\;(\mbox{GeV}^{-2})\le0.15$. Bold (lilac)
lines---total results, solid (black) lines---$O(1)$ contributions,
dashed (red) lines---$O(\alpha_{\rm s})$ contributions, dotted
(blue) lines---$O(\alpha_{\rm s}^2)$ contributions, dot-dashed
(green) lines---power contributions. Results from pole-mass OPE
(a) versus running mass OPE for renormalization scale
$\mu=2.5\;\mbox{GeV}$ (b), $\mu=3\;\mbox{GeV}$ (c), and
$\mu=4\;\mbox{GeV}$~(d).}\label{Plot:1a}\end{figure}

Let us define the scale $\tilde\mu$ by demanding
$M_b=\overline{m}_b(\tilde\mu).$ Using the $O(a^2)$ relation
between running and pole masses, we obtain numerically
$\tilde\mu\approx2.23\;\mbox{GeV}.$ The perturbative hierarchy of
the $\overline{\rm MS}$ expansion at this scale is \emph{worse\/}
than that of the pole-mass expansion: The $O(1)$ spectral
densities coincide,~whereas the $O(a)$ spectral density in the
$\overline{\rm MS}$ scheme receives a positive contribution
compared to the pole-mass scheme. For smaller scales, \emph{i.e.},
$\mu<\tilde\mu,$ the hierarchy of the $\overline{\rm MS}$
expansion becomes even worse~with decreasing $\mu$. For $\mu
>\tilde\mu$, first the hierarchy of the $\overline{\rm MS}$
expansion improves for increasing $\mu,$ see~Fig.~\ref{Plot:1a}.
However, as soon as the scale $\mu$ becomes sufficiently larger
than $\tilde\mu$, the ``induced'' contributions, which primarily
reflect the poor behaviour of the expansion of the pole mass in
terms of the running mass, overtake the ``genuine'' contributions.
This is illustrated by Fig.~\ref{Plot:1a}: at $\mu=4\;\mbox{GeV},$
the $O(1)$ contribution to the dual correlator rises steeply with
$\tau$, whereas the $O(a)$ contribution becomes negative in
order~to compensate the rise of the $O(1)$ contribution. Finally,
for large values of $\mu$ we observe a compensation between the
``induced'' contributions. One may expect that, in this case, the
accuracy of the expansion will deteriorate. This is reflected by a
strong scale dependence of the extracted values of $f_B$ and
$f_{B^*}$:~as $\mu$ rises, the $O(a)$ term ``undercompensates''
the rise of the $O(1)$ term for the $B$ meson and $f_B$ increases
with $\mu,$ whereas it overcompensates the rise of the $O(1)$ term
for the $B^*$ meson and $f_{B^*}$ decreases~with rising $\mu;$ as
a consequence, the ratio $f_{B^*}/f_B$ proves to be particularly
sensitive to the precise value of~$\mu.$

Now, returning to the pole-mass expansion for the $b$ quark, we
note that the hierarchy is not too~bad and can be easily improved
by switching to the running mass and choosing the scale $\mu$
slightly above~$\tilde \mu$. Furthermore, we may, for instance,
require that the $O(\alpha_{\rm s})$ contribution to the dual
correlator~remains positive in the working range of $\tau$. (For a
positive-definite dual correlator, it is not a strictly necessary
but, for obvious reasons, a highly welcome feature if each of the
perturbative contributions is positive.) In this case, we may
expect to arrive at reliable results by setting
$\mu=2.5$--$3\;\mbox{GeV}$ for the $B^*$ meson~and
$\mu=2.5$--$3.5\;\mbox{GeV}$ for the $B$ meson. An additional
important argument in favour of such~``low-$\mu$''~choice is that
the ratio $f_{B^*}/f_B$ proves to be definitely less than unity at
scales $\mu\approx m_b,$ whereas it emerges~close to unity for
$\mu=2.5$--$3\;\mbox{GeV},$ in full agreement with heavy-quark
expansion and hints from lattice~QCD.

The results for the $B^*$-meson decay constant shown in
Fig.~\ref{Plot:1a} have been found by employing,~for~the $b$-quark
mass, $\overline{m}_b(\overline{m}_b)=(4.18\pm 0.030)\;\mbox{GeV}$
\cite{pdg} and, for the other relevant OPE parameters, the values
\begin{align*}&\overline{m}_d(2\;\mbox{GeV})=(3.5\pm0.5)\;\mbox{MeV}\
,\qquad \overline{m}_s(2\;\mbox{GeV})=(95\pm5)\;\mbox{MeV}\
,\\&\alpha_{\rm s}(M_Z)=0.1184\pm0.0007\ ,\qquad\langle\bar
qq\rangle(2\;\mbox{GeV})=-[(269\pm17)\;\mbox{MeV}]^3\
,\\&\frac{\langle\bar ss\rangle(2\;\mbox{GeV})}{\langle\bar
qq\rangle(2\;\mbox{GeV})}=0.8\pm0.3\
,\qquad\left\langle\frac{\alpha_{\rm s}}{\pi}\,GG\right\rangle
=(0.024\pm0.012)\;\mbox{GeV}^4\ .\end{align*}
To compare the results obtained from, on the one hand, the
running-mass OPE and, on the other hand, the pole-mass OPE, we
recalculate the pole mass from the $O(a^2)$ relation between
$\overline{m}_b$ and $M_b,$ finding $M_b=4.8\;\mbox{GeV}.$ Merely
for illustrating the main features of the dual correlators, the
sum-rule estimates shown in Fig.~\ref{Plot:1a} are extracted for a
$\tau$-independent effective threshold $s_{\rm eff}=\mbox{const}.$
Its value in each~case is deduced by requiring maximal stability
of the extracted decay constant in the chosen~Borel window.

\section{\boldmath Extraction of observables such as decay
constants from QCD sum rules}The well-established procedures of
QCD sum rules pave a straightforward path to extract
observables:\begin{enumerate}\item Determine a reasonable Borel
window, that is, an interval of the Borel parameter $\tau$ defined
such that the OPE provides an accurate description of the exact
correlator: higher-order radiative and power corrections have to
be under control while, at the same time, the ground state
contributes ``sizeably'' to the correlator; our $\tau$ window for,
\emph{e.g.}, the $B$ meson reads
$0.05\lesssim\tau\;(\mbox{GeV}^{-2})\lesssim0.18.$\item Define and
apply an appropriate criterion for fixing the effective continuum
threshold $s_{\rm eff}(\tau).$ To this end, we employ an earlier
developed algorithm \cite{LMSETa,LMSETb,LMSETc,LMSETd} that allows
for a reliable extraction of the ground-state properties in
quantum mechanics and of the charmed-meson decay constants~in QCD.
We introduce the {\em dual invariant mass\/} $M_{\rm dual}$ and
the {\em dual decay constant\/} $f_{\rm dual}$ by defining
$$M_{\rm dual}^2(\tau)\equiv-\frac{{\rm d}}{{\rm
d}\tau}\log\Pi_{\rm dual}(\tau,s_{\rm eff}(\tau))\ ,\qquad f_{\rm
dual}^2(\tau)\equiv\frac{\exp\left(M_B^2\,\tau\right)}{M_B^4}\,
\Pi_{\rm dual}(\tau,s_{\rm eff}(\tau))\ .$$
The dual mass should reproduce the true ground-state mass $M_B;$
its deviation from $M_B$ quantifies the contamination of $\Pi_{\rm
dual}(\tau,s_{\rm eff}(\tau))$ by excited states. We determine the
behaviour of $s_{\rm eff}(\tau)$ by starting from a convenient
Ansatz for $s_{\rm eff}(\tau)$ and minimizing the deviation of the
predicted $M_{\rm dual}$ from the observed $M_B$ in the $\tau$
window by varying $s_{\rm eff}(\tau).$ Since we need to know the
behaviour of $s_{\rm eff}(\tau)$ only in the limited $\tau$ window
defined before, it suffices to consider merely polynomials in
$\tau$ (which Ansatz allows, of course, also for the case of
$s_{\rm eff}(\tau)$ being a $\tau$-independent constant):
$$s_{\rm eff}(\tau)=\sum\limits_{j=0}^ns_j^{(n)}\,\tau^{j}\
,\qquad\chi^2\equiv\frac{1}{N}\sum_{i=1}^N\left[M^2_{\rm
dual}(\tau_i)-M_B^2\right]^2\ .$$
\item With the variational result for $s_{\rm eff}(\tau)$ at hand,
$f_{\rm dual}(\tau)$ yields the decay-constant estimate~sought.
\end{enumerate}

As all outcomes for hadron observables extracted from QCD sum
rules do, also the predictions~for decay constants are sensitive
to the input values of all the parameters entering in one's
OPE---resulting in what we call, for clear reasons, their
OPE-related uncertainty---and to the particularities of the route
followed to get the effective threshold as a function of
$\tau$---contributing to their systematic uncertainty.
\begin{description}\item[OPE-related uncertainty:] By assuming a
Gaussian distribution for the numerical value of each of~the
parameters required as OPE input except for the renormalization
scales $\mu$ and $\nu,$ for which we assume uniform distributions
in the range $3\;\mbox{GeV}\leq\mu,\nu\leq6\;\mbox{GeV},$ we may
estimate the size of the OPE-related uncertainty by performing a
bootstrap analysis. The resulting distribution of decay
constants~proves~to be close to Gaussian shape. So, the quoted
OPE-related error should be understood as a Gaussian~error.
\item[Systematic uncertainty:] The systematic uncertainty is an
immediate consequence of the intrinsically limited accuracy of the
QCD sum-rule approach and thus poses, without surprise, a delicate
problem. Considering polynomial parameterizations of the effective
continuum threshold $s_{\rm eff}(\tau)$ for toy models within
quantum mechanics, we could demonstrate that the band of results
found from linear, quadratic, and cubic Ans\"atze for $s_{\rm
eff}(\tau)$ encompasses the true value of the decay constant.
Thus, the half-width~of this band should be regarded as a
realistic estimate for the systematic uncertainty of such an
extraction.\end{description}

\section{\boldmath Decay constants of pseudoscalar and vector
beauty mesons $B_{(s)}$ and $B^*_{(s)}$}
\begin{figure}[h]\centering\begin{tabular}{cc}
\includegraphics[width=6.687cm]{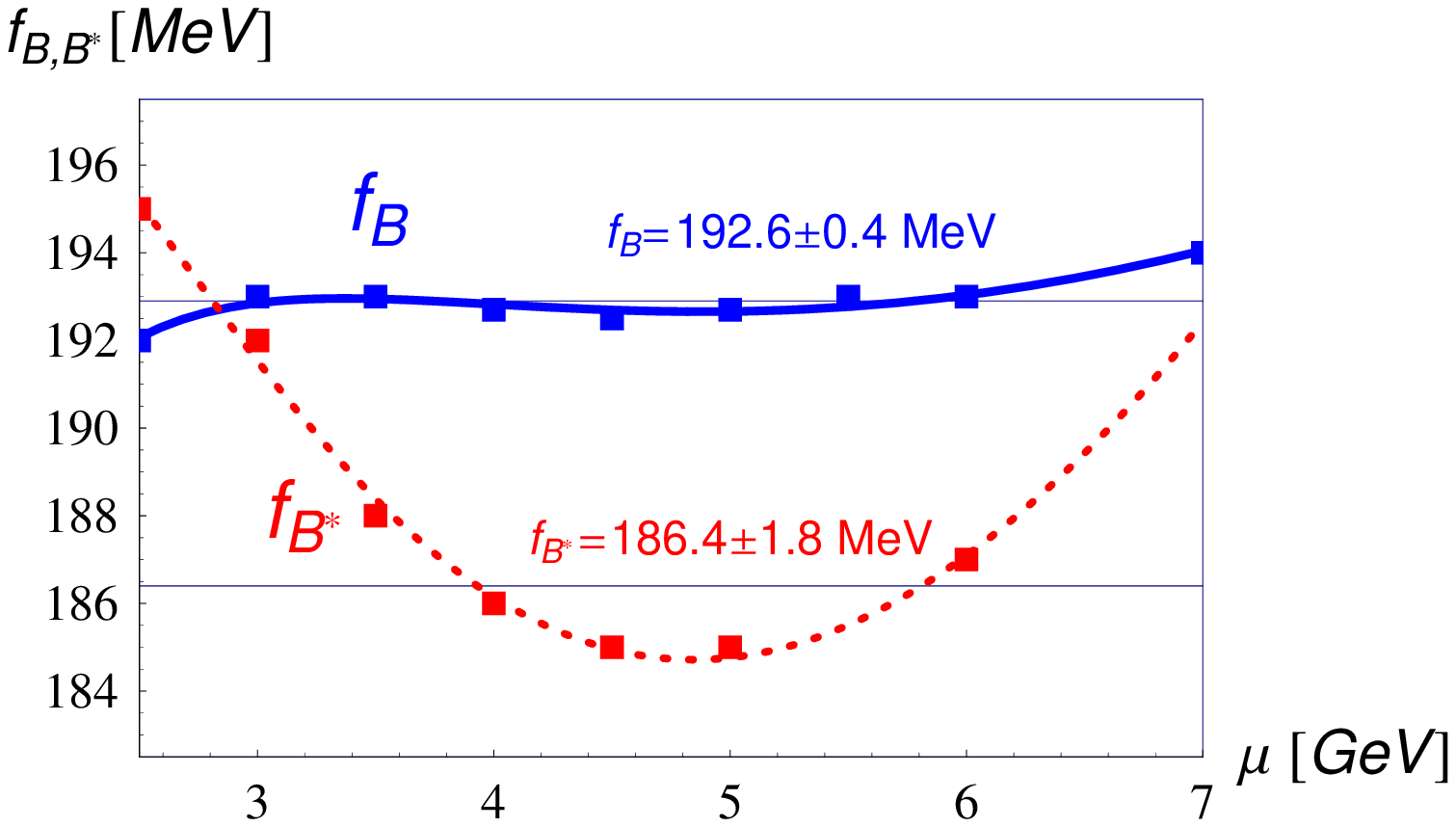}&
\includegraphics[width=6.687cm]{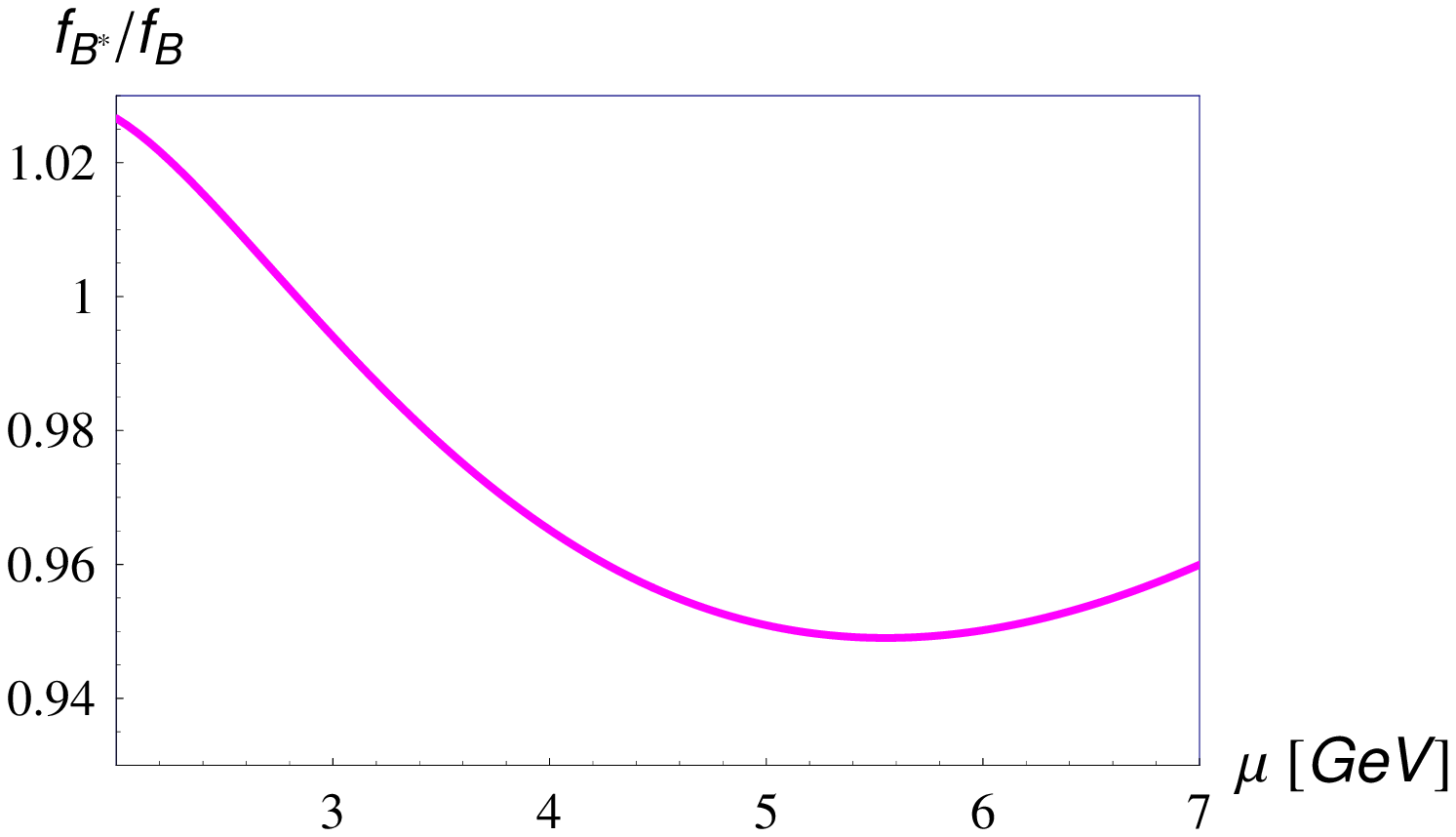}
\end{tabular}\caption{The $\mu$ dependence of the decay constants
$f_B$ and $f_{B^*}$ (left) and the ratio $f_{B^*}/f_B$ (right),
corresponding to central values of all OPE parameters except for
$\mu$ and to a quadratic Ansatz for the effective continuum
threshold.}\label{Fig:2}\end{figure}

The decay constants $f_B$
and $f_{B^*}$ emerging from application of our extraction
procedure exhibit a strong sensitivity to the value used for
$m_b\equiv\overline{m}_b(\overline{m}_b)$ (keeping fixed all other
parameters relevant for the OPE):\begin{itemize}\item The decay
constant $f_B^{\rm dual}(m_b,\mu)$ of the pseudoscalar meson $B$
behaves, as function of $m_b$ and $\mu,$~like
\begin{align*}&f_B^{\rm dual}(m_b,\mu=\mu^*)=192.6\;\mbox{MeV}
-13\;\mbox{MeV}\left(\frac{m_b-4.247\;\mbox{GeV}}{0.034\;\mbox{GeV}}\right),
\qquad\mu^*=5.59\;\mbox{GeV}\ ,\\&f_B^{\rm
dual}(m_b=4.247\;\mbox{GeV},\mu)=192.6\;\mbox{MeV}
\left(1-0.0015\log\frac{\mu}{\mu^*}+0.030\log^2\frac{\mu}{\mu^*}
+0.061\log^3\frac{\mu}{\mu^*}\right).\end{align*}
\item For the vector meson $B^*,$ our \emph{preliminary\/} results
for the dual decay constant $f_{B^*}^{\rm dual}(m_b,\mu)$ are
given~by
\begin{align*}&f_{B^*}^{\rm dual}(m_b,\mu=\mu^*)=186.4\;\mbox{MeV}
-10\;\mbox{MeV}\left(\frac{m_b-4.247\;\mbox{GeV}}{0.034\;\mbox{GeV}}\right),
\qquad\mu^*=5.82\;\mbox{GeV}\ ,\\&f_{B^*}^{\rm
dual}(m_b=4.247\;\mbox{GeV},\mu)=186.4\;\mbox{MeV}
\left(1+0.106\log\frac{\mu}{\mu^*}+0.337\log^2\frac{\mu}{\mu^*}
+0.173\log^3\frac{\mu}{\mu^*}\right).\end{align*}
\end{itemize}The scale $\mu^*$ is obtained by averaging over our
decay-constant predictions for $B$ and $B^*$ shown
in~Fig.~\ref{Fig:2}.

Our final sum-rule predictions for the decay constants under
consideration depend to a large extent on the chosen input value
of the $b$-quark mass $m_b$ and on the way one deals with the
dependence on~$\mu$:\begin{itemize}\item Assuming for $f_B$ and
$f_{B^*}$ a flat distribution in the interval
$\mu\in(3\;\mbox{GeV},6\;\mbox{GeV})$ and averaging over $\mu$ in
this range clearly yields for their ratio $f_{B^*}/f_B<1,$ largely
independent of the precise value of $m_b.$\item Using
$m_b=4.18\;\mbox{GeV}$ yields $f_B>210\;\mbox{MeV},$ in clear
tension with recent lattice-QCD results for~$f_B.$\end{itemize}In
Ref.~\cite{LMS_mb}, we noticed that requesting the sum-rule
prediction for $f_B$ to reproduce the lattice results requires the
substantially higher $b$-quark mass $m_b=4.247\;\mbox{GeV}.$ Now,
averaging for this $m_b$ our results found for a
\emph{quadratic\/} Ansatz for the effective threshold over $\mu$
in the range $\mu\in(3\;\mbox{GeV},6\;\mbox{GeV})$~yields
\begin{align*}f_B&=(192.6\pm1.6)\;\mbox{MeV}\ ,&\quad
f_{B^*}&=(186.4\pm3.2)\;\mbox{MeV}\ ,\\f_{B_s}&=(231.0\pm
1.8)\;\mbox{MeV}\ ,&\quad f_{B_s^*}&=(215.2\pm3.0)\;\mbox{MeV}\
,\end{align*}
where the uncertainties quoted above are merely those brought
about by the dependence on the scale~$\mu.$

\section{Cursory summary of observations, conclusions, and outlook}
\begin{enumerate}\item For beauty mesons, a strong correlation
between $m_b$ and the sum-rule result for $f_B$ was observed:
$$\frac{\delta f_B}{f_B}\approx-8\,\frac{\delta m_b}{m_b}\ .$$
Combining our sum-rule analysis with the latest results for $f_B$
and $f_{B_s}$ from lattice QCD~implies
$$m_b=\left(4.247\pm0.027_{\rm OPE}\pm0.011_{\rm
syst}\pm0.018_{\rm exp}\right){\rm GeV}\ .$$
\item Whereas the decay constants of charmed mesons, $f_D,$
$f_{D_s},$ $f_{D^*},$ and $f_{D_s^*}$, obtained from QCD~sum rules
\cite{LMSDa,LMSDb,LMSD*} turned out to be practically independent
of the particular choice of the scale $\mu$, in the beauty sector
the situation is different: the decay constants of bottom mesons,
particularly of vector bottom mesons, are very sensitive to the
precise value of the scale $\mu$. Averaging over~the range
$3<\mu\;(\mbox{GeV})<6,$ we get the non-strange and strange
bottom-meson decay-constant~ratios
$$\frac{f_{B^*}}{f_B}=0.923\pm0.059\ ,\qquad
\frac{f_{B_s^*}}{f_{B_s}}=0.932\pm0.047\ ;$$
here, the above uncertainties incorporate the OPE-related
uncertainties as well as the systematic uncertainties estimated
along the course of our algorithm. If, however, one relies on
calculations performed for low $\mu$ scales,
$2.5<\mu\;(\mbox{GeV})<3.5,$ then, to a good accuracy, one finds
$f_{B^*}/f_B\approx1.$\end{enumerate}The unpleasant dependence of
the QCD sum-rule predictions for the beauty-meson decay
constants~on the scale $\mu$ requires further detailed study in
order to acquire better control over the related uncertainty.

\end{document}